\documentclass[twoside]{article}
\usepackage{qic,epsfig}
\begin{document}
\setlength{\textheight}{8.0truein}    

\runninghead{ }
            { }

\normalsize\textlineskip
\thispagestyle{empty}
\setcounter{page}{1}

\copyrightheading{0}{0}{2003}{000--000}

\vspace*{0.88truein}

\alphfootnote

\fpage{1}

\centerline{\bf
Some Attacks On Quantum-based Cryptographic Protocols}
\vspace*{0.37truein} \centerline{\footnotesize
Hoi-Kwong Lo \footnote{ Email: hklo@comm.utoronto.ca  }}
\vspace*{0.015truein} \centerline{\footnotesize\it Center for
Quantum Information and Quantum Control} \baselineskip=10pt
\centerline{\footnotesize\it Dept. of Electrical \& Computer
Engineering \& Dept. of Physics} \baselineskip=10pt
\centerline{\footnotesize\it  University of Toronto, 10 King's
College Road } \baselineskip=10pt \centerline{\footnotesize\it
Toronto, Ontario, CANADA, M5S 3G4.
} \vspace*{10pt} \centerline{\footnotesize Tsz-Mei Ko \footnote{
   Email: tszmei@us.ibm.com }} \vspace*{0.015truein}
\centerline{\footnotesize\it IBM East Fishkill} \baselineskip=10pt
\centerline{\footnotesize\it
 Department of Physical Synthesis Bldg 334, Rm 2K17/424 }
\baselineskip=10pt \centerline{\footnotesize\it
 2070 Route 52
Hopewell Junction, NY 12533, USA }
 \vspace*{0.225truein} \publisher{(received date)}{(revised
date)}

\vspace*{0.21truein}

\abstracts{
Quantum-based cryptographic protocols are often said to enjoy
security guaranteed by the fundamental laws of physics. However,
even carefully designed quantum-based cryptographic schemes may be
susceptible to subtle attacks that are outside the original
design. As an example, we give attacks against a recently proposed
``secure communication using mesoscopic coherent states'', which
employs mesoscopic states, rather than single-photon states. Our
attacks can be used either as a known-plaintext attack or in the
case where the plaintext has not been randomized. One of our
attacks requires beamsplitters and the replacement of a lossy
channel by a lossless one. It is successful provided that the
original loss in the channel is so big that Eve can obtain $2^k$
copies of what Bob receives, where $k$ is the length of the seed
key pre-shared by Alice and Bob. In addition, substantial
improvements over such an exhaustive key search attack can be
made, whenever a key is reused. Furthermore, we remark that, under
the same assumption of a known or non-random plaintext, Grover's
exhaustive key search attack can be applied directly to "secure
communication using mesoscopic coherent states", whenever the
channel loss is more than 50 percent. Therefore, as far as
information-theoretic security is concerned, optically amplified
signals necessarily degrade the security of the proposed scheme,
when the plaintext is known or non-random. Our attacks apply even
if the mesoscopic scheme is used only for key generation with a
subsequent use of the key for one-time-pad encryption. Studying
those attacks can help us to better define the risk models and
parameter spaces in which quantum-based cryptographic schemes can
operate securely. Finally, we remark that our attacks do not
affect standard protocols such as Bennett-Brassard BB84 protocol
or Bennett B92 protocol, which rely on single-photon signals.}{}{}

\vspace*{10pt}

\keywords{Quantum Cryptography, Quantum Key Distribution,
Unconditional Security} \vspace*{3pt} \communicate{to be filled by
the Editorial}

\vspace*{1pt}\textlineskip    
\section{Introduction}
\label{S:Intro}

``Cryptographers do not sleep well.'' In conventional
cryptography, it is well-known that cryptographic protocols may
fail to achieve the designed goals. It is unfortunate that those
design flaws are often not fully appreciated until it is too late.
A well-known example is Queen Mary of Scotland whose usage of an
insecure encryption scheme to communicate with her conspirators in
a plot against Queen Elizabeth led directly to her execution.
Another example is the breaking of the legendary German Enigma
code in World War II by the Allied code-breakers. The Allied
code-breaking effort contributed significantly to the outcome of
the war and involved a) a number of innovative mathematical
discoveries, b) the construction of large-scale code-breaking
machines (which were precursors to modern computers), c) the
exploitation of a number of mistakes/weaknesses in the
implementation details of the German and, finally, d) the
capturing of an Enigma machine. The first lesson in cryptography
is: never under-estimate the effort and ingenuity that your
adversaries are willing to spend on breaking your codes. Given
that cryptography can be a matter of life and death, there is no
surprise that cryptographers are generally paranoid.

Quantum mechanics has made a remarkable entry in cryptography.
Unlike conventional cryptographic schemes that are based on
computational assumptions, it is often said that quantum-based
cryptographic schemes have their security guaranteed by the
fundamental laws of physics. Various cryptographic protocols based
on quantum mechanics have been proposed in the literature. The
best-known protocol is quantum key distribution (QKD). The idea of
QKD was first published by Bennett and Brassard in 1984
\cite{bb84}. It took more than ten years for people to prove
rigorously that QKD can, in principle, provide perfect security
guaranteed by quantum mechanics
\cite{mayersqkd,qkd,biham,benor,shorpre,kaoshi,gl}. Even in the
case of imperfect devices, unconditionally secure QKD is still
possible under rigorous theoretical models\cite{ilm,gllp}.
Provided that we are willing to be conservative cryptographers, we
can adjust the parameters in our experiment so that we have
unconditional security proven in a precise physical model, whose
validity can then be battle-tested. The battle-testing will
hopefully lead to a refinement of our physical model and so on and
so forth. Provided that an eavesdropper cannot break our precise
physical model at the time of transmission of quantum signals, we
can obtain unconditional security.

It is, however, very important to distinguish unconditional
security offered by precise physical models from ad hoc security
derived from handwaving arguments. Whenever the security of a
protocol relies on handwaving arguments, there is serious danger
that hidden loopholes may exist in a protocol. The bottom line is
that it pays to make our physical model as precise and explicit as
possible and it pays to battle-test our physical model in a real
QKD system. Indeed, technological advances are intrinsically
difficult to predict. What is not realistic today may become
realistic tomorrow. Worse still, given that security agencies such
as the NSA has numerous mathematicians and physicists working hard
on quantum technologies and not all of their results will be
published, it is rather risky to assume that ad hoc security will
survive their future attacks.

Moreover, even carefully designed cryptographic schemes may be
susceptible to subtle attacks that are outside the original
design. It would, therefore, be interesting to study those subtle
loopholes in existing schemes. In this paper, we will examine new
attacks against existing quantum-based cryptographic protocols.
One of our attacks is a beamsplitting attack against a ''secure
communication using mesoscopic coherent states'' \cite{yuen},
which employs mesoscopic signals, rather than single-photon
signals. Our attack applies when the plaintext is known, or more
generally, non-random. Furthermore, we remark that, under similar
assumptions, Grover's exhaustive key search attack can be applied
directly to "secure communication using mesoscopic coherent
states", whenever the channel loss is more than 50 percent.

\section{Beamsplitting plus low loss channel attack}
\noindent

The idea of using mesoscopic states for quantum key distribution
was first proposed by Bennett and Wiesner \cite{bwpatent} in 1996.
Recently, there has been much interest in the subject of
quantum-based cryptographic schemes with coherent states. In
particular, Barbosa {\it et al.} \cite{yuen} have described an
experimental implementation of one particular scheme which employs
mesoscopic coherent states. Such a scheme can be optically
amplified and can open up new avenues for secure communications at
high speeds in fiber-optic or free-space channels. Just imagine
transmitting quantum signals through optical amplifier and one
could obtain long distance secure communication. It is, thus,
important to understand the limitations of those schemes.

Let us now describe Barbosa et al.'s scheme \cite{yuen} in more
detail. In \cite{yuen}, the two users, Alice and Bob, initially
share a short $k$-bit seed key, $K$. Using a (deterministic)
pseudo-random number generator, each of them can expand their seed
key, $K$, which is a classical string, into a longer running key,
$K'$, which is also a classical $M$-ary string. Say, $K' = ( K'_0,
K'_1, \cdots , K'_{n-1})$ where $K'_i \in \{0 , \cdots, M-1 \}$.
Suppose Alice would like to transmit some classical data $X= (
X_0, X_1, \cdots, X_{n-1} )$, which is an $n$-bit binary string,
to Bob. Then, they use the running key $K'$ to encrypt the data
$X$. The transmitted signal is a sequence of quantum states. In
order to describe what states are transmitted, we need to
introduce some standard states. More specifically, consider $M$
states $ | \alpha_0 ( \cos \theta_l + i \sin \theta_l ) \rangle $,
where $\theta_l = 2 \pi l / M $ and $0 \leq l \leq M -1 $. Note
that, for each even $M$, they form $M/2$ pairs $\{ l, l + M/2 \}$.
Each pair forms a signal set. The two signals of each pair are
almost orthogonal for a large $\alpha_0$. In fact, the inner
product between any two signals in a pair is $exp ( - 2 | \alpha_0
|^2 )$. Barbosa et al.'s scheme \cite{yuen} considers the scenario
when $M \gg \alpha_0$ so that the $M$ states are not (almost)
distinguishable. Each component, say $K'_i$, of the running key is
used to specify which one of the $M$ bases is used for encoding
the data bit, $X_i$. Indeed, the transmitted state is of the form
$ |Y_0 \rangle \otimes | Y_1 \rangle \otimes \cdots  \otimes |
Y_{n-1} \rangle $ where $  Y_i = \alpha_0 (  \cos \theta_l + i
\sin \theta_l )$ where $l = K'_i$ if $X_i =0$ and $l = K'_i + M/2$
if $X_i =1$. If an eavesdropper has no information on the running
key $K'$, then it might appear that she cannot do much better than
random guessing.

Instead of considering a ciphertext only attack as originally
studied in \cite{yuen}, we consider either a known-plaintext
attack or the case where the plaintext has not been randomized.
Known-plaintext attacks were employed successfully against both
the Germans and the Japanese during the Second World War. We will
discuss a simple theoretical attack that can break such a scheme
with a known or non-random plaintext. It involves beamsplitters
and the replacement of the original high loss channel by a low
loss channel. Our observation is the following: By replacing a
high loss channel by a low loss channel, an eavesdropper obtains
many copies of the output state received by the receiver, Bob. In
principle, she can use beamsplitters to split the signal, sending
one copy to Bob and keeping many copies herself. Now, Eve is at
liberty to perform an exhaustive key search to break the
enciphering scheme, in analogy with exhaustive key search attack
in conventional cryptography. [Note that such an exhaustive search
attack is possible if the plaintext is known or, more generally,
not random.] Useful improvements over such an exhaustive search
attack will also be presented later in this paper.

Let us begin with an exhaustive search attack. More concretely,
suppose the loss of the channel is such that only a fraction $ 1/
( t+ 1)$ of the quantum signal reaches Bob. Then, by replacing a
lossy channel by a lossless one and intercepting the residual
signal, Eve can obtain $t$ copies of the signal received by Bob.
In the spirit of exhaustive key search, for each copy of the
signal, Eve can try to decode it with a specific trial value for a
seed key. In more detail, She picks a trial value $K =
K_{trial_i}$ for the seed key and runs a pseudo-random generator
to obtain a running key $K'_{trial_i}$. With such a running key in
her hand, she can then measure a copy of the signal to try to
determine the value of $X$. Note that if she has guessed the value
of the seed key correctly, she will most likely obtain the correct
value of $X$ from her measurement.

Our attack applies to two cases---as a known-plaintext attack or
the case where the plaintext has not been randomized. Consider
case one: known-plaintext attack. As in standard cryptography, we
are given $P_1, C_1= E_k(P_1), P_2, C_2 =E_k (P_2), \cdots, P_i,
C_i=E_k(P_i)$. The eavesdropper's job is to deduce $k$ or an
algorithm that will infer $P_{i +1} $ from $C_{i+1} = E_k(
P_{i+1})$. Known-plaintext attacks allow the eavesdropper to
verify that she has guessed the correct encryption key. In case
two, the plaintext, $X$, is not known. However, we assume that it
is {\it not} randomized. Suppose $X$ is a text in English, there
is a lot of redundancy in the message. Eve will be able to verify
that the message is in English. On the other hand, if her trial
value $K = K_{trial_i}$ for the seed key is incorrect, then it is
highly unlikely that she will find that the decoded message is a
meaningful English text. In summary, she will be able to tell
whether a trial seed key is correct or not simply by seeing
whether the decoded message is meaningful or not. Now, suppose the
number $t$ of copies in Eve's hand satisfies $t \geq 2^k$ where
$k$ is the number of bit of the seed key. Then, Eve could have
tried all possible combinations of the seed key. Therefore, Eve
will have succeeded in breaking the security of the scheme
completely. This conclusion applies to both cases (case
one---known-plaintext attack and case two---where the plaintext is
unknown, but not randomized).

Some remarks are in order. First, our attack may be outside the
original design of the protocol. The original design of the
protocol \cite{yuen} seems to focus on ciphertext only attacks.

Second, our attack appears to be feasible with current or near
future technology: Since installed fibers may go through many
intermediate switches, they may not take the shortest path between
the sender and the receiver. Therefore, it is not unrealistic to
imagine that the eavesdropper may re-route the original optical
path to a shorter one, thus effectively reducing the loss in the
channel. Besides, low loss fibers are under active development in
various laboratories in the world \cite{rutgers} and they give
much lower loss than standard Telecom fibers through which QKD
experiments are usually performed. Therefore, Eve may apply those
low loss fibers for eavesdropping purposes.

For an eavesdropping attack that involves re-routing the light
through a shorter path, it may lead to a different time delay in
the channel. Therefore, Alice and Bob may try to defeat it by
testing the transmission time. However, such a testing by Alice
and Bob may be defeated by a technologically advanced Eve: Given
that there have been successful "stopping light" experiments
\cite{hau,lukin}, we think it may not be unrealistic to believe
that Eve might be able to reproduce the same transmission time by
delaying the light pulses only temporarily in the near future.
Therefore, detecting the transmission time may not be sufficient
to defeat our attack.

Third, when Alice is sending a long message to Bob, Eve may not
even need that many copies of the signal to decode it. To
illustrate this point, let us consider, for simplicity the case
when the plaintext is a message of the form, $P_1 P_2 \cdots P_r$
where $P_i$ is $s$-bit long and $rs =n$. In fact, we allow Alice
to divide up her transmission of the message into $r$ different
time windows. For instance, $P_1$ might be transmitted on the
first day (or hour or minute), $P_2$ on the second day, ..., $P_r$
on the $r$-th day. We assume that Eve has no long-term quantum
memory (but is able to use beamsplitters and measure mesoscopic
signals). Once again, let us assume that $P_i$ is a text in
English or is a known plaintext. Suppose the channel is lossy and
only a fraction $ 1 / (t +1)$ of the signal reaches Bob. Then,
effectively, Eve has $t$ copies of all the $r$ messages, $X_i$ in
some encrypted form in different time windows. Altogether, Eve has
$rt$ encrypted messages. Now, Eve can try a different trial seed
key value $K_{trial_i}$ for each of the $rt$ messages. Therefore,
provided $rt \geq 2^k$, Eve will succeed in an exhaustive key
search and find the encryption key, $k$. From that point on, she
will be able to infer any future message $P_{r+1}$ from its
ciphertext $E_k(P_{r+1})$. Note that given a fixed transmission
fraction $ 1 / (t +1)$, for a sufficiently large $r$, the scheme
will become insecure. This sets a fundamental limit on how many
times Alice and Bob can expand a key before the scheme becomes
insecure against a ``beamsplitting plus exhaustive key search
attack''.

Fourth, Eve can substantially increase the power of an exhaustive
key search attack by changing the value of her trial seed key
mid-way in the decoding process. In fact, for each encrypted copy
of a message, $X_i$, she may first try a trial seed key value,
$K_{trial_i}$ on the first few bits of the ciphertext. She checks
whether the decrypted message is a meaningful text in English. If
it is, she will continue the decryption process for a few more
bits and checks again and so on and so forth. On the other hand,
whenever it becomes clear that the decrypted part of the message
is not a meaningful text in English, Eve can change her trial seed
key value from  $K_{trial_i}$ to another value, say $
K_{trial_j}$. She decrypts a few bits and sees if the resulting
text makes sense. If it does not, she can change the trial value
of the trial seed key again. The process is done until she is
fairly confident that she has found the right trial seed value.

We remark that one countermeasure that Alice and Bob can use to
circumvent this particular improved attack is to perform double
encryption. The message $P_i$ should be encrypted by a standard
classical cipher like AES (Advanced Encryption Standard) before
passing through a quantum-based encryption device.

Fifth, if Eve has a quantum computer, then she can launch a much
more powerful attack. Indeed, she will be able to decode the
message with high probability whenever the transmission ratio of
the channel is less than 50 percent. The idea is that Eve can
perform a Grover's \cite{grover} database search on the encrypted
message to see whether she has found a correct key. See
\cite{grover,brassard,bound} for details. Even though the
ciphertext is a quantum state, we remark here that, in the context
of known-plaintext attacks, such a quantum exhaustive key search
method by Grover can be applied directly to \cite{yuen}.
Furthermore, only {\em one} plaintext-ciphertext pair will be
sufficient for performing Grover's search. Therefore, it is
sufficient for Eve to have a single copy of the signal. This
implies that the scheme is, in principle, insecure against a
known-plain quantum computing attack, whenever the transmission
ratio is less than 50 percent. Why does a Grover's search work? We
remark that we are considering the asymptotic case where Alice and
Bob are using a pseudo-random number generator to expand the seed
key by a very large factor. In this case, the quantum codewords
after the encoding corresponding to the various possible seed key
values will asymptotically become {\it orthogonal} to each other.
Therefore, there exists a measurement by Eve that will distinguish
between those codewords, thus disclosing the value of the seed
key.

A variant of the proposal in \cite{yuen} is to use the mesoscopic
quantum state for key generation only and subsequently use the
generated key for one-time-pad encryption. Here, we will show that
our attacks apply even to this variant. First of all, let us
describe the variant. As before, Alice expands a short seed key,
$K$, that she shares with Bob into a long running key, $K'$, using
a classical pseudo-random number generator. Now, Alice generates a
random number, $R$, and uses the mesoscopic quantum encryption
method of Barbosa {\it et al} \cite{yuen} to send the encrypted
version of $R$, i.e., $E_{K'} (R)$, to Bob. [Here, $E_{K'} (R)$ is
a quantum state and $K'$, the long running key, is used as an
encryption key.] Bob decodes the message and recovers $R$. Note
that Alice and Bob now share a common key $R$. Now, suppose Alice
would like to send a message, $M$, to Bob. She uses the random
number $R$ as a one-time-pad and sends $C = R \bigoplus M$ to Bob.
Bob uses $R$ to decode the message and recover $M$. Now consider
the security of this variant of a proposal in \cite{yuen}. First,
consider a known-plaintext attack. Eve is given a set of old
plaintext together with their cipher-texts $M_1, C_1= R_1
\bigoplus M_1, M_2, C_2 =R_2 \bigoplus M_2, \cdots, M_i, C_i=R_i
\bigoplus M_i$. Here, the long string of random number $R $ is a
concatenation of $R_1 R_2 \cdots R_i \cdots$. Note, however, that
such a known plaintext attack means that Eve can obtain $R_1 = M_1
\bigoplus C_1$, etc, which are now the plaintext used in the
mesoscopic key generation scheme. In other words, a
known-plaintext attack on the variant can be reduced to a
known-plaintext attack on the original protocol. Similarly, a
non-random plaintext in this variant can be reduced to a
non-random plaintext in the original protocol. For this reason,
our attacks (based on known plaintext or non-random plaintext)
apply directly to this variant of the protocol \cite{yuen} too.

\section{Concluding Remarks}
\noindent In summary, attacks against quantum-based cryptographic
schemes are discussed in this paper. Our attacks can be used to
defeat a communication protocol based on mesoscopic coherent
states. Such a scheme can be optically amplified. Our attack
applies if a plaintext is known or, more generally, not random. We
remark again that known-plaintext attacks were successfully
launched against both the Germans and the Japanese in World War
II. Our attack involves replacing a high loss channel by a low
loss channel and using beamsplitters to steal a component of the
signal. Therefore, our attack can, to some extent, be implemented
with current technology. Once Eve has copies of the quantum
signal, she can perform exhaustive key search on her copies to
find out the value of the key. Moreover, she can improve
substantially over a simple exhaustive key search by exploiting
redundancies in the plaintext, unless double encryption has been
performed. In fact, given any transmission ratio $ 1 / (t +1)$ of
the channel, the scheme will become insecure whenever a seed key
is expanded more than $r$ times where $rt \geq 2^k$.

Furthermore, if Eve has a quantum computer, she can apply Grover's
database search algorithm to attack the scheme. Such a
quantum-computational attack will work whenever the loss of the
channel is over 50 percent.

Our attacks apply not only to the original protocol in
\cite{yuen}, but also to a variant where the protocol in
\cite{yuen} is used only for key generation and subsequently
one-time-pad encryption is employed, using the generated key.

Our attacks show that there are fundamental limits on the amount
of key expansion that can be securely achieved using mesoscopic
states. In future investigations, it will be interesting to
quantify this fundamental limit more precisely.

We remark that it will be interesting to look into attacks that
require technologies intermediate between simple beamsplitters and
a large-scale quantum computer.

As a general remark, breaking cryptographic systems is as
important as building them. Our attacks are rather simple but
powerful. They highlight that, just like their classical
counterparts, the study of the security of quantum-based
cryptographic schemes is a very slippery subject. It is important
to make security assumptions as explicit as possible.

In summary, our attacks highlight the fact that even carefully
designed cryptographic schemes may still be susceptible to attacks
outside the original design. Understanding those attacks will thus
be helpful to improving the security of those schemes, for
example, by better defining the parameter space or risk models in
which those schemes are secure.

Finally, we remark that our attacks do not apply to
BB84\cite{bb84} or other standard QKD schemes where the quantum
signals are strictly microscopic in the sense that there is (on
average) at most one copy of the signal available.

\nonumsection{Acknowledgements} \noindent We thank helpful
discussions with colleagues including G. Barbosa, O. Hirota, M.
Koashi, D. Leung, N. L\"{u}tkenhaus, D. Mayers, T. Mor, J.
Preskill and H. P. Yuen. Useful comments from an anonymous referee
are gratefully acknowledged. H.-K. L. thanks CFI, CIPI, CRC
program, NSERC, OIT, and PREA for financial support.

\end{document}